\documentclass[12pt]{iopart}
\usepackage[english]{babel}
\usepackage{iopams}
\usepackage{graphicx}

\newcommand{\ket}[1]{|#1\rangle}
\newcommand{\bra}[1]{\langle#1|}

\begin{document}

\title{Robustness of Heisenberg-limited interferometry with balanced Fock states}


\author{D Meiser and M J Holland}
\address{JILA, National Institute of Standards and Technology,
and University of Colorado, Boulder, CO 80309-0440, USA}
\eads{dmeiser@jila.colorado.edu}

\begin{abstract}
Interferometry with Heisenberg limited phase resolution may play an important
role in the next generation of atomic clocks, gravitational wave detectors, and
in quantum information science. For experimental implementations the robustness
of the phase resolution is crucial since any experimental realization will be
subject to imperfections. In this article we study the robustness of phase
reconstruction with two number states as input subject to fluctuations in the
state preparation. We find that the phase resolution is insensitive to
fluctuations in the total number of particles and robust against noise in the
number difference at the input. The phase resolution depends on the uncertainty
in the number difference in a universal way that has a clear physical
interpretation: Fundamental noise due to the Heisenberg limit and noise due to
state preparation imperfection contribute essentially independently to the
total uncertainty in the phase. For number difference uncertainties less than
one the first noise source is dominant and the phase resolution is essentially
Heisenberg limited. For number difference uncertainties greater than one the
noise due to state preparation imperfection is dominant and the phase
resolution deteriorates linearly with the number difference uncertainty.
\end{abstract}

\section{Introduction}

The precision with which phase shifts can be determined in an interferometer is
ultimately limited by shot noise. The shot noise originates from the
discreteness of the interfering particles \footnote{For conciseness we refer to
the interfering objects as particles with the understanding that they could be
realized as actual particles or field quanta such as photons.}. For
interferometry with independent particles, each particle interferes only with
itself. For $N$ independent particles the precision is $\sqrt{N}$ times better
than the single particle precision, a result which follows from application of
the central limit theorem. This scaling of the phase precision with the number
of particles,
$
\Delta \phi_{\rm SQL}\sim 1/\sqrt{N},
$
is called the standard quantum limit. 

In some applications an increase in the number of interfering particles is
difficult or impractical.  In these cases it is worth trying to make better use
of the available resources. Interferometry with phase resolution better than
the standard quantum limit is relevant for gravitational wave detectors
\cite{Gustafson:LIGO,Kimble:SqueezingGWProposal,McKenzie:SqueezingGWDetection}.
The best atomic clocks employing ${\rm Cs}$
\cite{PhysRevLett.82.4619,SalomonCsClock}, trapped ions
\cite{oskay:020801,T.Rosenband03282008}, and earth-alkaline atoms such as
${}^{87}{\rm Sr}$ in optical lattices
\cite{Takamoto:LatticeClock,ludlow:033003,LeTargat1,Boyd2,Ludlow2} are also
operating very close to the standard quantum limit. In the next generation
of clocks the precision could be further increased by using interferometry with
resolution better than the standard-quantum-limit. 

Phase resolution better than the standard quantum limit can be achieved with
entangled many particle states.  A variety of such input states have been
proposed. For interferometry with photons, squeezed states of light have been
suggested. Bollinger and Wineland have
proposed spin squeezed states for ions and other massive particles
\cite{PhysRevA.46.R6797}. Maximally entangled states have been proposed theoretically \cite{PhysRevA.54.R4649,VittorioGiovannetti11192004}
and their improved phase resolution has been demonstrated experimentally
\cite{PhysRevLett.86.5870}. Holland and Burnett have suggested to use Fock
states in each of the input ports and they have shown that these states yield
Heisenberg limited phase resolution \cite{HollandBurnett}.  States
with improved phase resolution have been systematically studied in
\cite{Summy:PhaseOptimizedStates,PhysRevLett.85.5098,PhysRevA.54.4564,Hermann:Interferometry,huang:250406}.

For linear interferometers the resolution achievable with $N$ particles is
fundamentally limited to a scaling $\Delta\phi_{\rm H.L.}\sim 1/N$, called the Heisenberg limit.

So far, most studies of interferometry with resolution better than the standard
quantum limit have concentrated on ideal realizations in which the initial
state can be realized with perfect fidelity, no dephasing or decoherence
happens during the evolution through the interferometer, and the final state is
detected with ideal detectors. For the maximally entangled Schr{\"o}dinger cat
state, the effect of dephasing during the evolution through the interferometer
has been studied by Huelga \etal \cite{HuelgaInterferometryDissipation}. The
effect of decorrelation on the phase resolution for interferometry with
balanced Fock states has been studied by Kim \etal \cite{PhysRevA.57.4004} and
recently Dorner \etal \cite{Dorner:OptimalPhaseEstimation} have constructed
generalizations of maximally entangled states that are more robust against
photon loss inside the interferometer.

The goal of this article is to study the robustness of the dual Fock state
against imperfections in the creation of the initial state. We find that this
state is insensitive to fluctuations in the total number of particles fed into
the interferometer. The phase resolution is robust against fluctuations in the
number difference. If the fluctuations in the number difference are of order
one, the phase resolution is still essentially Heisenberg limited. For
increasing imbalance uncertainty, the phase resolution degrades linearly
and becomes standard quantum limited as the imbalance uncertainty approaches
the partition noise.  These findings add a whole class of states to the arsenal
of states that are candidates for phase resolution better than the standard
quantum limit. This class of states can be thought of as the incoherent cousin
of the conventional squeezed states. The states are incoherent mixtures of
number difference states with a probability distribution that is narrower than
the partition noise.

\section{\label{model} Model}

\begin{figure}
\hspace{3cm}\includegraphics[width=10cm]{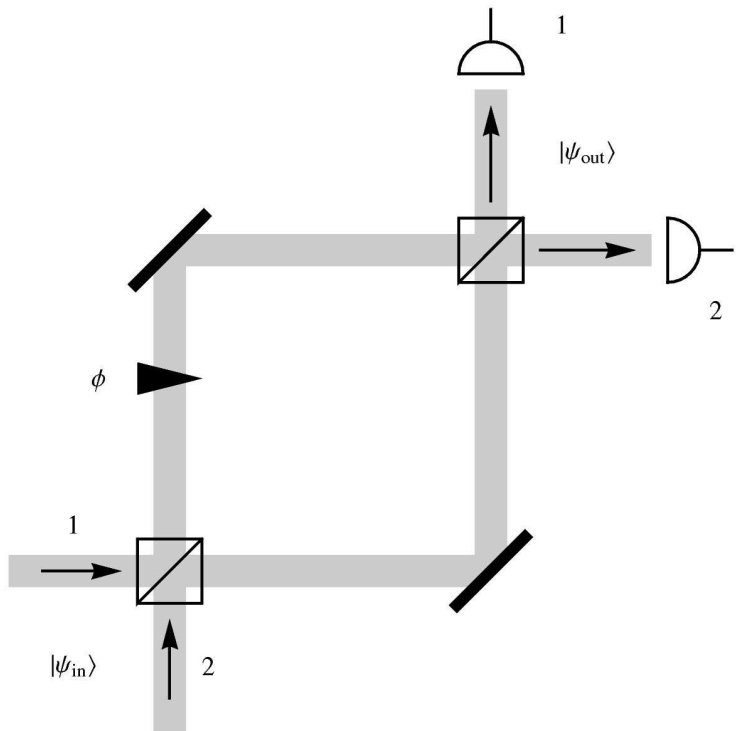}
\caption{Schematic of a Mach-Zehnder interferometer. Particles prepared in
state $\ket{\psi_{\rm in}}$ pass through a beam splitter. The particles going
through one arm acquire a phase shift $\phi$ relative to the other arm. After
recombination at the second beam splitter the particles coming out of ports one
and two are detected.} \label{InterferometerSchematic}
\end{figure}

For the remainder of this article we consider the elementary Mach-Zehnder 
interferometer depicted in figure \ref{InterferometerSchematic}. Particles
prepared in state $\ket{\psi_{\rm in}}$ pass through a beam splitter. The
component passing through one of the arms experiences a phase shift $\phi$
relative to the other arm. The two modes are recombined at a second beam
splitter and detectors measure the numbers of particles coming out in ports one
and two.

We describe the particles passing through the interferometer as effective two
level systems with the two states corresponding to the two arms of the
interferometer. Mathematically we describe these two level systems using the
angular momentum formalism. This formalism as it applies to
interferometry has been described in detail in several articles, see for
instance \cite{Hermann:Interferometry,PhysRevA.57.4004,
Yurke:AngularMomentumInterferometer}.
In brief, each particle is identified with a fictitious spin-1/2 system with
spin up and down corresponding to the particle being in port one and two,
respectively. A collection of $N$ particles is then described by combining
the $N$ spins into a total angular momentum.

The $z$-component $\hat J_z$ of the total angular momentum describes the number
imbalance between the ports of the interferometer and the square of the length
of the angular momentum vector $\hat J^2$ is related to the total number of
interfering particles. We choose the $\hat J_z$ and $\hat J^2$ eigenstates
$\ket{J,m}$ with\[
\hat J_z \ket{J,m}=m\ket{J,m}
\]
and \[
\hat J^2\ket{J,m}=J(J+1)\ket{J,M}
\]
and $J=N/2$ as basis states. The perfectly balanced Fock state suggested by Holland and
Burnett is the state $\ket{J,0}$ 
\footnote{For simplicity we restrict ourselves to even total numbers of
particles. The case of odd numbers of particles is qualitatively
identical.}. The interferometer transformation that relates the output state to
the input state is
\[
\ket{\psi_{\rm out}}=e^{i\phi\hat J_y}\ket{\psi_{\rm in}}.
\]
From here on we specialize to the case of an interferometer 
with relative phase shift $\phi=0$. For the balanced input states that we consider
below this choice of working point corresponds to the null fringe were equal
numbers of particles are detected in both output ports of the interferometer.

\subsection{State preparation}

In an ideal experiment it would be possible to deterministically create the
state $\ket{J,0}$ with fixed $J$. However, for systems with large numbers of
particles, where the precision gain would be most dramatic, this perfect
situation cannot be achieved in general. In practice, states $\ket{J,m}$ are
created stochastically with a probability distribution $\mathcal{P}_{J,m}$.
Typically it is impossible to determine which $J$ and $m$ were realized due to
finite detector sensitivity. This means that one has to reconstruct the phase
based on incomplete knowledge of the initial state.

It has been shown by Uys and Meystre \cite{Hermann:Interferometry} that the
phase resolution for states $\ket{J,m}$ depends very little on $m$ as long
as $|m|\ll J$. \footnote{In fact, these authors have shown that the phase
resolution is nearly Heisenberg limited almost all the way to $|m|\sim J$.} We
can therefore assume, without loss of generality, that the mean of the
distribution $\mathcal{P}_{J,m}$ is centered at $m=0$ in accordance with the
original proposal of Holland and Burnett \cite{HollandBurnett}. For the working
point $\phi=0$ this is also the most interesting case from an experimental
perspective, where it is advantageous to detect approximately equal numbers of
particles in both detectors due to finite detector efficiency. Modeling the
distribution $\mathcal{P}_{J,m}$ as a Gaussian distribution in $J$ and $m$ we
can characterize the state preparation by three parameters: the average total
angular momentum $J$ and the standard deviations $\Delta
J=\sqrt{\sum_{J^\prime}\mathcal{P}_{J^\prime}(J^\prime-J)^2}$ and $\Delta
m=\sqrt{\sum_m\mathcal{P}_m m^2}$ of the distributions of the total angular
momentum $J$, $\mathcal{P}_J=\sum_m \mathcal{P}_{J,m}$, and of the number
imbalance $m$, $\mathcal{P}_m=\sum_J\mathcal{P}_{J,m}$. From now on we refer to
$\Delta m$ as the state preparation imperfection or just imperfection for
brevity. The initial state is then a statistical
mixture,
\begin{equation}
\hat \rho=\sum_{J,m} {\mathcal P}_{J,m} \ket{J,m}\bra{J,m}.
    \label{initialstate}
\end{equation}

\subsection{Phase reconstruction: Bayesian inference and likelihood function}

It is well known that reconstructing the phase shift from a
measurement record is a non-trivial task if the input state of the
interferometer is the balanced Fock state. For that state, the mean number
imbalance $\langle \hat J_z\rangle$ is zero for all phase shifts and hence
cannot be used as a signal. The variance $(\Delta \hat J_z)^2$ varies
sinusoidally with the phase, but the signal-to-noise ratio of that observable is
of order one. Thus a large number of measurements are necessary in order to
infer the phase shift. It can be shown that the same holds true for all higher
moments of $\hat J_z$ as well. In order to achieve that state's potential phase
resolution, one has to use a Bayesian reconstruction scheme instead
\cite{HollandBurnett}. This scheme has the additional advantage that it yields
the full probability distribution of the phase conditioned on a measurement
record instead of {\it e.g.} the first few moments. Furthermore, it immediately yields
a prescription for how to find the phase in an actual experiment. 

The basic idea of Bayesian inference is to update the probability distribution
of the phase according to
\begin{equation}
P(\phi|m)=\frac{P(\phi)P(m|\phi)}{\mathcal{N}},
\label{BayseRule}
\end{equation}
with $\mathcal{N}$ a normalization constant. Note that we use
the symbol $\mathcal{N}$ to denote different normalization constants throughout
this article whose precise value is irrelevant. The prior phase distribution
$P(\phi)$ contains all the knowledge available about the phase before the first
measurement. To provide an unbiased analysis of the phase measurement we assume
that nothing is known about the phase initially except that it lies in the
interval $[-\pi/2,\pi/2]$.  \footnote{This choice of the prior phase
distribution makes this phase inference method essentially a maximum likelihood
estimation.} This latter
restriction removes the irrelevant phase ambiguity 
where phases that differ from each other by an integer multiple
of $\pi$ cannot be distinguished. The prior phase distribution is then
\begin{equation}
P(\phi)=
\left\{\begin{array}{rl}
\pi^{-1}, & |\phi|\le \pi/2\\
0, & |\phi|>\pi/2.\end{array}\right.
\end{equation}

$P(m|\phi)$ is the conditional probability for a number difference $m$ to occur
given that the phase shift was $\phi$. The denominator normalizes the
probability distribution $P(\phi|m)$.

Equation (\ref{BayseRule}) can be iterated: After a measurement of
the number difference $m_1$ the knowledge about the phase is $P(\phi|m_1)$. The
phase distribution after a second measurement is then obtained from equation
(\ref{BayseRule}) by using $P(\phi|m_1)$ as the prior. For a measurement record
$\{ m_1,m_2,\ldots\}$ one finds the phase distribution
\begin{equation}
P(\phi|m_1,m_2,\ldots)=\frac{P(m_1|\phi)P(m_2|\phi)\cdot \cdots}{\mathcal{N}},
\end{equation}
 The sharpness of this
distribution after a certain number of measurements is a figure of merit of the
phase resolution obtainable with a given input state.

In the limit of a large number of measurements $n$ the various possible
measurement outcomes $m$ occur $n P(m|\theta)$ times, where $\theta$ is the
actual phase shift. The phase distribution then becomes
\begin{eqnarray}
P(\phi|\theta)&={\mathcal{N}}^{-1}\prod_{m=-J}^J P(m|\phi)^{n P(m|\theta)}\\
		&={\mathcal{N}}^{-1}F(\phi|\theta)^n,
    \label{asymptoticDistribution}
\end{eqnarray}
with
\begin{equation}
F(\phi|\theta)=\prod_{m=-J}^J P(m|\phi)^{P(m|\theta)}.
\label{likelihoodF}
\end{equation}
The function $F(\phi|\theta)$ is the asymptotic likelihood function of the phase. It
fully characterizes the expected phase resolution in the limit of large $n$.

\section{\label{results}Results}

\subsection{\label{conditionalProbsSection} Conditional probabilities $P_{\hat \rho}(m|\phi)$}

From equation (\ref{BayseRule}) it is clear that the conditional probabilities
$P(m|\phi)$ are of central importance for the phase resolution. The sharper the
distribution $P(m|\phi)$ is as a function of $\phi$, the more one learns
about the phase if the number difference $m$ is detected. However it is not
sufficient for a state to have sharply peaked $P(m|\phi)$ distributions in
order to be useful for interferometry. Measurement outcomes $m$ with such sharp
distributions also must be likely to occur. This latter requirement makes the
phase resolution typically dependent on the working point of the
interferometer, {\it i.e.} the precision depends on the phase that one wishes to
measure. 

The conditional probabilities $P(m|\phi)$ are calculated by sending the initial state $\hat \rho$ of equation (\ref{initialstate}) through the interferometer and projecting the final state on the subspace with a given $m$,
\begin{equation}
P_{\hat \rho}(m|\phi)={\rm Tr}\left( e^{i \phi \hat J_y}\hat \rho e^{-i\phi\hat J_y}\hat \Pi_m\right),
    \label{condprobsdef}
\end{equation}
where $\hat \Pi_m$ is the projector on the $m$ subspace,
\[
\hat \Pi_m =\sum_J |J,m\rangle\langle J,m|.
\]
The conditional probabilities for a well defined initial $J$ and $m$, $P_{\ket{J,m_0}\langle J,m_0|}(m|\phi)$, behave as
\begin{equation}
P_{\ket{J,m_0}\langle J,m_0|}(m|\phi)=J_{m_0-m}^2(J \phi)
\end{equation}
for phase shifts small enough so that the number differences $m$ near the poles
of the Bloch sphere are unlikely. In this equation $J_{m_0-m}$ is the Bessel function of the first kind of order $m_0-m$.

\begin{figure}
\hspace{1cm}\includegraphics[width=15cm]{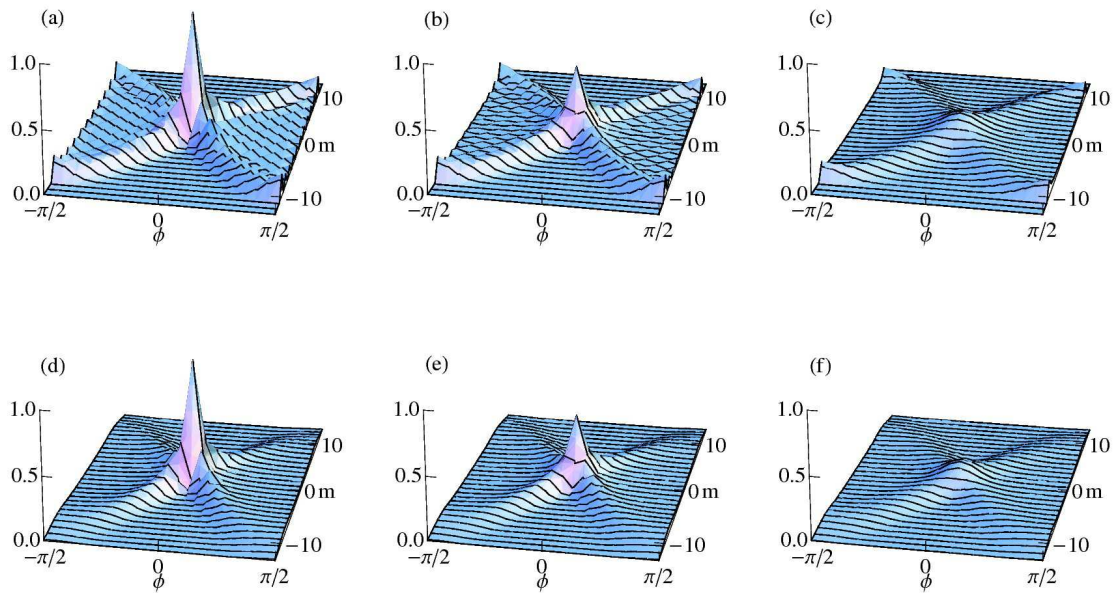}
\caption{Conditional probabilities $P_{\hat \rho}(m|\phi)$ for initial states
$\hat \rho$ for different levels of uncertainties $\Delta J$ and $\Delta m$ for
$J=10$. In the first row there is no uncertainty in the total particle number, 
$\Delta J=0$, and the number difference uncertainties are (a)$\Delta m=0$ (perfect state preparation),
(b) $\Delta m=1$, and (c) $\Delta m=3$. With increasing state preparation
imperfection $\Delta m$ the central peak of the distribution becomes broader
leading to reduced phase resolution. Panels (d-f) show the probability
distributions for the same state preparation imperfections as the panels in the
first row but with $\Delta J=3$. These panels demonstrate that additional
uncertainty in $J$ does not affect the central peak of the probability
distributions. Rather such noise affects the wings of the distributions for
phase shifts $\phi\gtrsim 1/\Delta J$.}
\label{CondProbs}
\end{figure}

In Figure \ref{CondProbs} we show the conditional probabilities for various
initial states for an average of 20 particles, i.e.  $J=10$.  As a consequence
of the Bessel function like behavior, the conditional probability $P(0|\phi)$
for perfect state preparation in figure \ref{CondProbs}(a) is sharply peaked at
$\phi=0$ with a width of order $\sim 1/J$. This sharp peak gives rise to
Heisenberg limited phase resolution for the perfectly balanced Fock state.  For
larger phase shifts the conditional probabilities oscillate with frequency $J$
and with an envelope falling off as $2/(\pi J \phi)$.

\begin{figure}
\hspace{1cm}
\includegraphics[width=6.5cm]{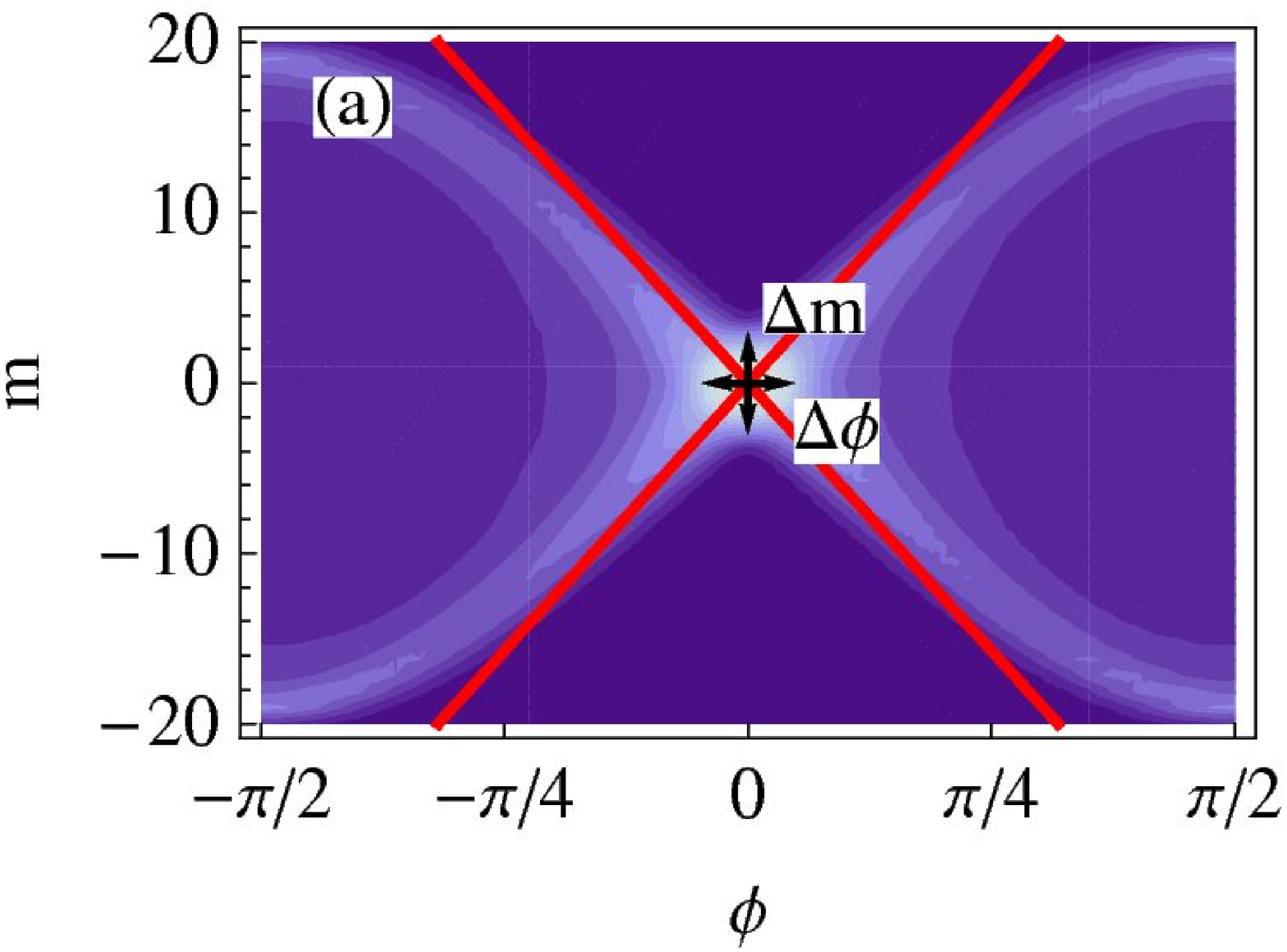}
\hspace{1cm}
\includegraphics[width=6.5cm]{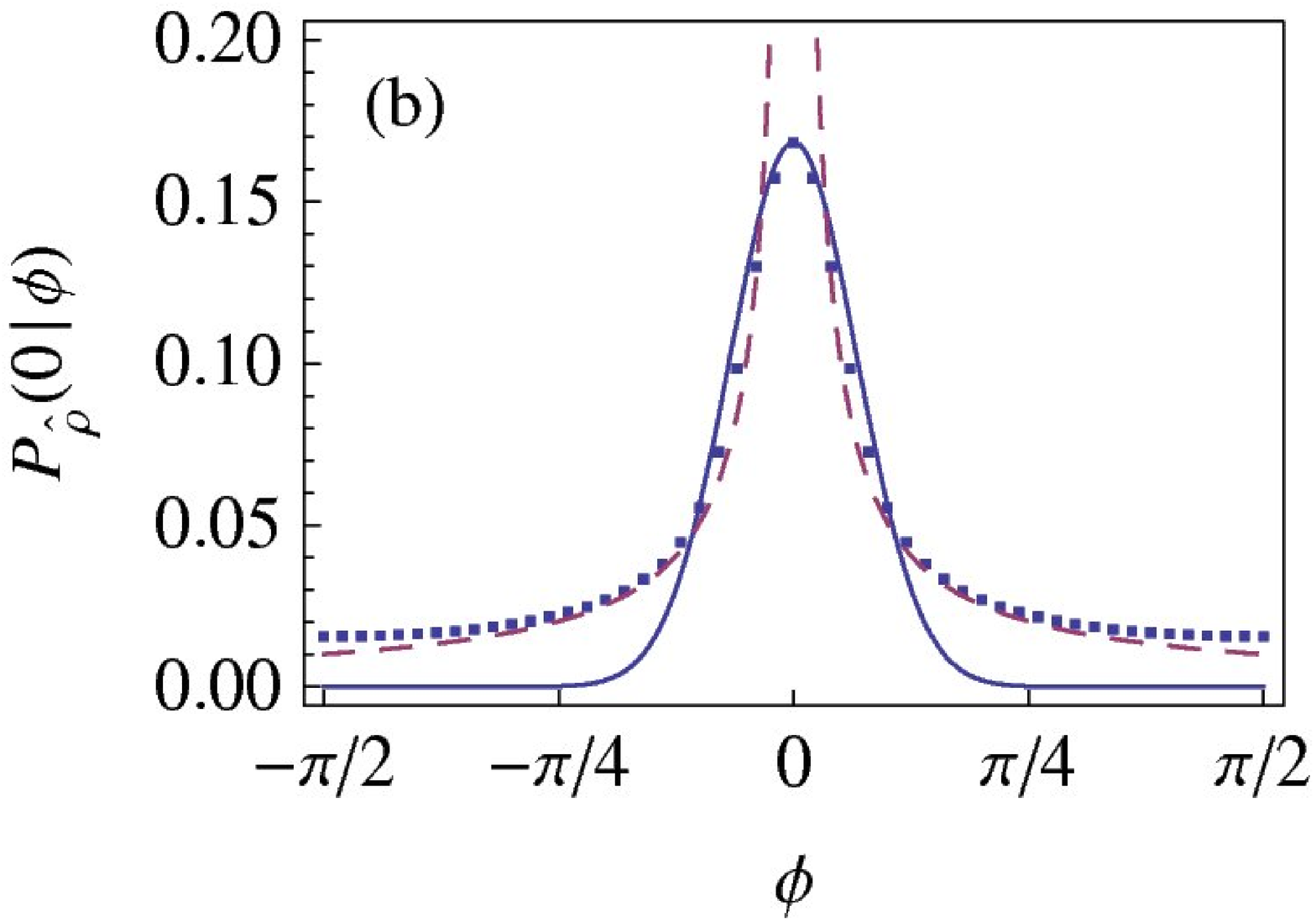}
\caption{Figure (a) illustrates the construction of the estimate of the width
    of $P_{\hat \rho}(0|\phi)$ for $J=20$, $\Delta J=0$  and $\Delta m=3$.
	The red lines indicate curves $m=\pm J \phi$. The two arrows indicate
	the widths of Gaussian distributions along $m$ and $\phi$. Figure (b)
	shows the agreement between the central portion of
	$P_{\hat\rho}(0|\phi)$ and a Gaussian with width $\Delta \phi=\Delta
	m/J$ normalized to the same height (blue solid line) for parameters
	identical to (a). For phase shifts larger than $1/J$ good agreement is
	found between $P_{\hat \rho}(0|\phi)$ and $(\pi J |\phi|)^{-1}$
	(purple dashed line) which follows from the asymptotic behavior of the Bessel functions.}
	\label{dmdphiEstimate}
\end{figure}

Figures \ref{CondProbs}(b,c) show that the width of the peak of the conditional
probabilities increases with increasing state preparation imperfection $\Delta
m$. Qualitatively, it is clear that this increased width leads to a decreasing
phase resolution. The width of the central peak can be estimated by the
construction illustrated in figure \ref{dmdphiEstimate}(a). For phase shifts
$|\phi|\ll \pi/2$ the maxima of the distribution $P(m|\phi)$ follow the red
lines given by $m=\pm J \phi$. Mathematically, this follows from properties of the Bessel-functions which describe the conditional probabilities well for $|\phi|\ll \pi/2$. Qualitatively, the scaling of $m$ with $\phi$ can be understood by noting that the number difference has to increase from zero to $|m|=J$ for a phase shift of $\pm\pi/2$. These lines translate the Gaussian
distribution $P(m|0)$ of $m$ with width $\Delta m$ into a Gaussian distribution of the
phase with width
\begin{equation}
\Delta \phi=\frac{\Delta m}{J}.
\label{dphiestimate}
\end{equation}
This Gaussian is compared with the exact probability distribution $P(0|\phi)$
in figure \ref{dmdphiEstimate}(b). The central portion of that probability
distribution is well described by the Gaussian. The wings of the conditional
probability distributions can be found from the asymptotic behavior of the
Bessel functions. Averaging several Bessel functions leads to the tails falling
off as
\begin{equation}
P_{\hat \rho}(m|\phi)\approx \frac{1}{\pi J \phi}.
\end{equation}
This is also illustrated in figure \ref{dmdphiEstimate}(b). Note that in a phase
reconstruction experiment the wings get more and more suppressed with
increasing number of measurements and eventually the phase distribution is
entirely determined by the approximately Gaussian central peak.

The figures \ref{CondProbs}(d-f) show the same conditional probabilities as in
the panels (a-c) but with $\Delta J=3$. As a consequence of the addition of
Bessel functions that asymptotically oscillate with different frequencies the
oscillations in the tails wash out for phase shifts $|\phi|\gtrsim 1/\Delta J$ in
figure \ref{CondProbs}(d). Near the poles $|m|\sim J$ of the Bloch sphere the
distributions look different compared to case with a sharp total number of
particles, $\Delta J=0$, because the total angular momentum now has a spread.
Most importantly however, the central Gaussian peak is unaffected by the noise
in $J$. This is true for noise up to $\Delta J\sim J$. As a consequence, the
phase reconstruction is largely insensitive to this type of noise. Physically,
this is because the measurement is based on detection of number differences in
which the total number of particles cancels out. Only when the phase shift is
so large that almost all particles exit the interferometer in one port does the
total number of particles play a role. We have numerically confirmed this
independence of the phase resolution of $\Delta J$. That allows us to consider
the limit $\Delta J\to 0$ from now on without loss of generality.

\subsection{\label{finitenmeas}Simulation of phase reconstruction: Finite numbers of measurements}

To illustrate one of the key new features of phase reconstruction with
imperfect state preparation compared to the ideal case we simulate an actual
phase reconstruction experiment. As in the numerical examples in the previous
section we consider a total particle number of $N=20$ and the state preparation
imperfection is $\Delta m=1$. 

To simulate the phase reconstruction we draw a sequence of measurement results
$m$ at random with probabilities given by $P_{\hat \rho}(m|\phi=0)$. Using the
known conditional probabilities $P_{\hat \rho}(m|\phi)$ we then successively
update the phase distribution by means of Bayes' formula. The evolution of the
phase distribution with successive measurements is not deterministic due to the
stochastic process with which the measurement results are determined.

\begin{figure}
\hspace{1cm}\includegraphics[width=6.5cm]{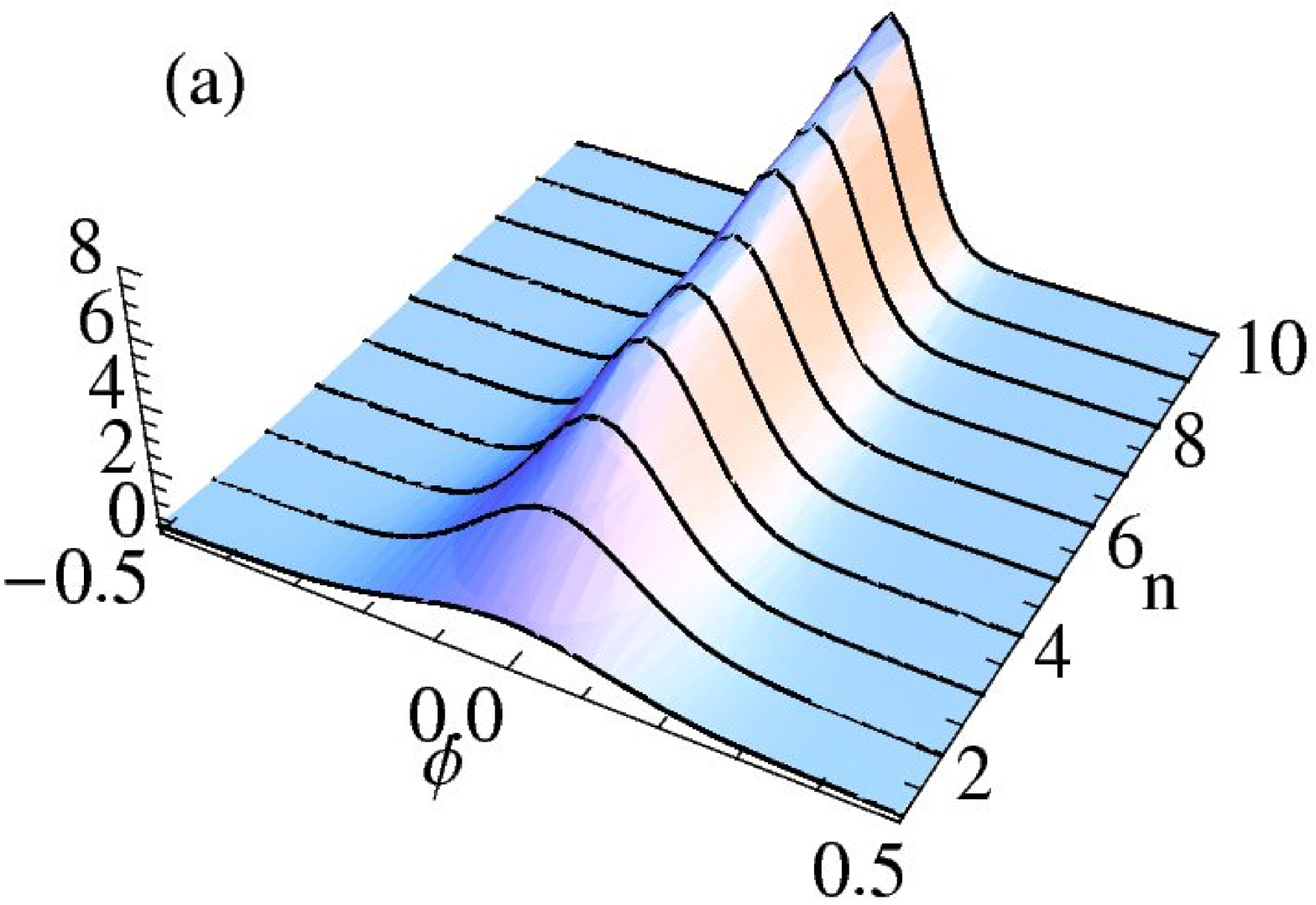}\hspace{1cm}
\includegraphics[width=6.5cm]{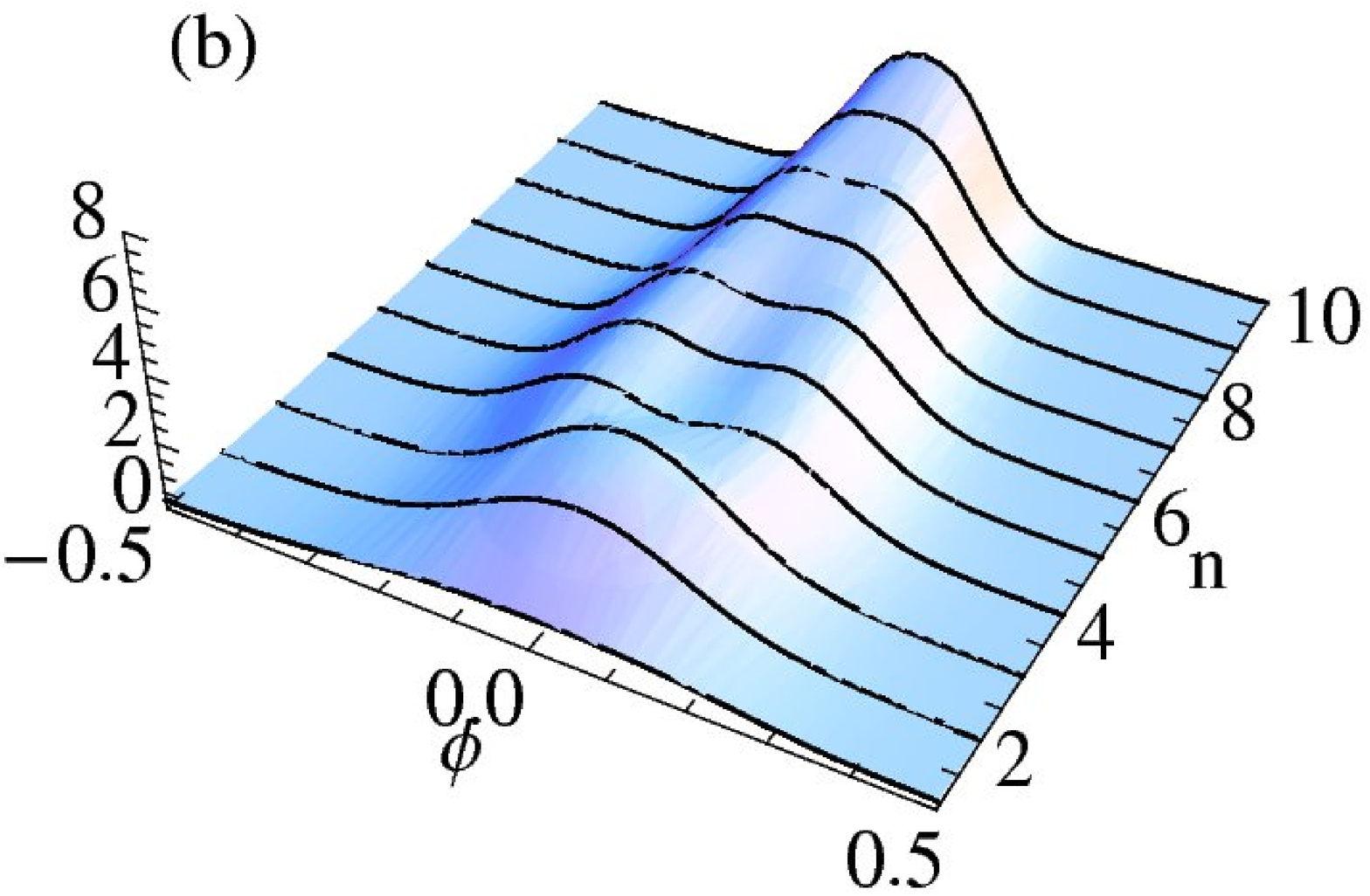}
\caption{Two different runs of numerical phase reconstruction simulations. The
figures show the phase distributions $P(\phi|m_1,m_2,\ldots,m_n)$ as a function of
the length of the measurement record, $n$. Panel (a) shows a reconstruction run
that is similar to what one finds for interferometry with perfect state
preparation. In panel (b) an unfavorable measurement outcome of $m=-2$ in the fourth
measurement leads to a wrong inference of a non-zero phase shift. This
inaccuracy is remedied by further measurements.}
\label{ExampleReconstructions}
\end{figure}

Two representative examples of such phase reconstruction simulations are shown
in figure (\ref{ExampleReconstructions}). The first example is reminiscent of phase
reconstruction with perfect state preparation. The phase distribution is
centered at the correct phase shift $\phi=0$ and the width of the distribution
shrinks monotonically with increasing number of measurements.

The second example illustrates a complication that can happen with imperfect
state preparation. In the fourth measurement of that sequence a number
difference of $m=-2$ is detected. For the state preparation imperfection $\Delta m=1$ this
measurement outcome is relatively unlikely to occur for zero phase shift and
the Bayesian updating wrongly attributes the measurement result to a non-zero
phase shift. The phase distribution becomes wider and is peaked at non-zero
phase shift. Thus this unfavorable measurement outcome reduces the phase
resolution and, even worse, it leads to an inaccurate phase inference. Further
measurements correct this measurement and in the limit of large numbers of
measurements the phase inference is accurate.

General quantitative statements on how many measurements are necessary in order
to rule out such inaccuracies for a certain level of state preparation
imperfection are difficult to make. In practice it is probably easiest to
simply run numerical simulations of the phase reconstruction in order to
empirically decide how many measurements should be made.

\subsection{Asymptotic phase resolution}

Asymptotically, the phase resolution $\Delta \phi$ improves as $\sim
1/\sqrt{n}$, where $n$ is the number of measurements. A figure of merit that is
independent of the number of measurements and characterizes the
suitability of an input state for phase measurements is therefore
\begin{equation}
\Delta \phi_\infty =\lim_{n\to \infty}\sqrt{n}\Delta\phi(n),
    \label{asymptoticscaling}
\end{equation}
where $\Delta \phi(n)$ is the standard deviation of the phase distribution
after $n$ measurements. We can calculate $\Delta \phi_\infty$ through 
the variance of the phase distribution $P(\phi|\theta)$ given in equation
(\ref{asymptoticDistribution}) with an $n$ that is large enough so that the
scaling equation (\ref{asymptoticscaling}) holds. For the results that follow
we have used $n=5$. This asymptotic phase distribution is devoid of the
statistical fluctuations encountered in the previous section for small numbers
of measurements.

\begin{figure}
\hspace{2.5cm}\includegraphics[width=12cm]{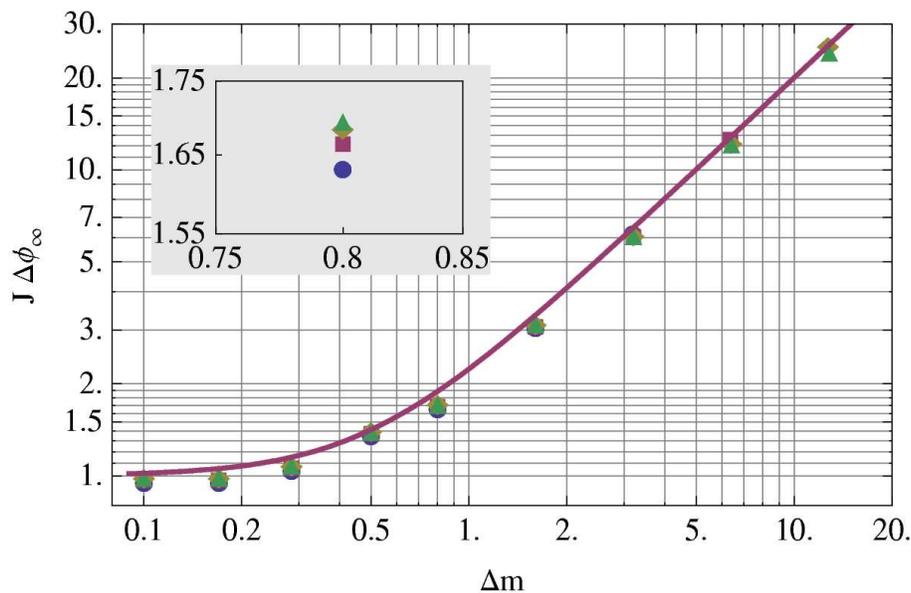}
\caption{Numerically determined asymptotic phase resolution $\Delta
\phi_\infty$ as a function of the state preparation imperfection $\Delta m$.
The figure shows the resolution rescaled by $J^{-1}$ for $J=10$ (blue discs),
$J=20$ (purple squares), $J=40$ (yellow diamonds), and $J=80$ (green
triangles). The resolutions rescaled this way fall on a universal curve in
agreement with formula (\ref{dphiinftyest}). The inset shows the data points
for $\Delta m=0.8$ at a higher resolution. The resolution described by equation
(\ref{dphiinftyest}) with $\alpha=2$ is shown by the purple line.}
\label{phi0figure} \end{figure}
 
We have numerically calculated the asymptotic phase resolution $\Delta
\phi_\infty$ as a function of $\Delta m$ for different values of $J$. The
results are shown in figure \ref{phi0figure}. In that figure we have rescaled
the phase resolution by $J^{-1}$. Remarkably, the resolutions for different $J$
rescaled this way all collapse onto a single universal curve. For nearly
perfect state preparation the resolution saturates at the Heisenberg limit,
\begin{equation}
\Delta \phi_\infty J\to 1,\qquad {\rm for}\quad \Delta m<1.
\end{equation}
For $\Delta m> 1$, $\Delta \phi_\infty$ increases linearly with $\Delta
m$.

The universal behavior of $J\Delta \phi_\infty$ can be understood by treating
the fundamental $1/J$ noise due to the Heisenberg limit and the noise due to
the state preparation imperfection as two independent noise sources. The total
noise is then obtained by adding these noise sources in quadrature. According
to the construction explained in figure \ref{dmdphiEstimate} the noise due to
the state preparation imperfection is given by $\alpha \Delta m/J$, where
$\alpha$ is a constant of order one. $\alpha$ describes by how much
$F(\phi|\theta=0)$  is broadened compared to the Gaussian in figure
\ref{dmdphiEstimate}(b) due to the admixture of results $m\neq 0$. Adding then the Heisenberg limit noise and the state preparation noise in quadrature we find
\begin{equation}
\Delta \phi_\infty=\sqrt{J^{-2}+(\alpha\Delta m/J)^2}.
\label{dphiinftyest}
\end{equation}
This scaling of the phase resolution with $J$ and state preparation
imperfection $\Delta m$ is the central result of this article. We find
excellent agreement between equation (\ref{dphiinftyest}) and the numerically
determined phase resolutions for $\alpha=2$, as illustrated by the purple line
in figure \ref{phi0figure}.

\subsection{Phase reconstruction with unknown $\Delta m$}

In a typical experiment one will not precisely know the initial number
difference uncertainty $\Delta m$. Especially in light of the numerical
simulations for finite numbers of measurements presented in section
\ref{finitenmeas} it is important to understand the dependence of the phase
inference on the choice of an estimate for $\Delta m$. In particular one needs to know under what
circumstances the phase distribution is asymptotically accurate. Or, phrased
differently, how should one estimate $\Delta m$ so that unfavorable measurement
results like in figure \ref{ExampleReconstructions}(b) are eventually corrected
by a large number of measurements.
\begin{figure}
\hspace{3cm}\includegraphics[width=12cm]{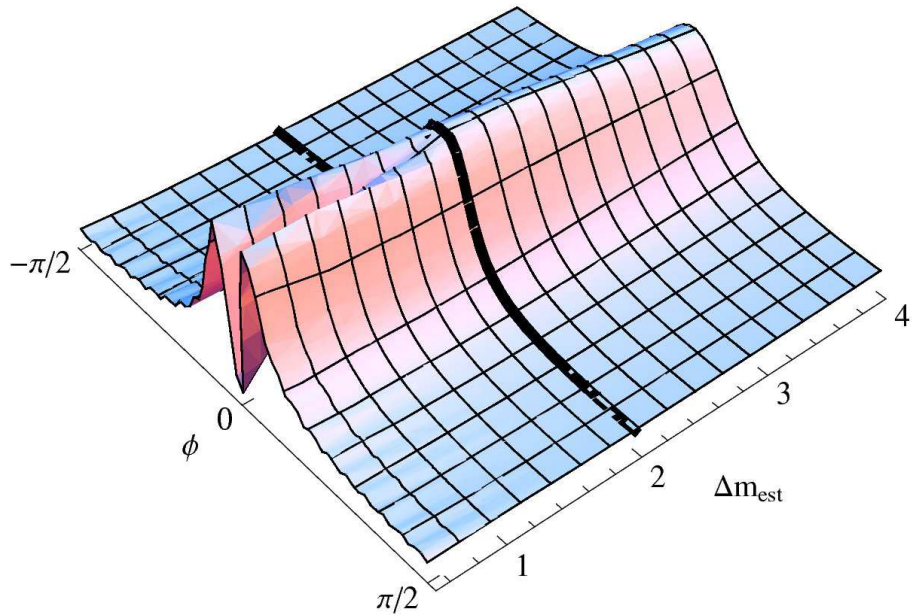}
\caption{Likelihood function $F(\phi|\theta=0;\Delta m_{\rm est},\Delta m)$ of the phase as a function of the estimate $\Delta m_{\rm est}$ for the number difference uncertainty for $J=20$. The actual number difference uncertainty $\Delta m=2$ is marked by the black line. For too optimistic estimates $\Delta m_{\rm est}<\Delta m$ the likelihood function becomes bimodal and the asymptotic phase distribution is inaccurate. For conservative estimates $\Delta m_{\rm est}>\Delta m$ the asymptotic phase distribution is accurate but broader than for $\Delta m_{\rm est}=\Delta m$.}
\label{Fphidmest}
\end{figure}

This problem can be studied by looking at the likelihood function
$F(\phi|\theta)$ of equation (\ref{likelihoodF}) as a function of the estimate
$\Delta m_{\rm est}$ for the number difference uncertainty. Basically one
pretends that the number difference fluctuations are $\Delta m_{\rm est}$ and
calculates the corresponding conditional probabilities $P_{\hat \rho(\Delta
m_{\rm est})}(m|\phi)$ using the initial state $\hat \rho(\Delta m_{\rm est})$
with imperfection $\Delta m_{\rm est}$. But in an actual experiment each
measurement result occurs with probability $P_{\hat \rho(\Delta m)}(m|\phi)$
where the true fluctuations $\Delta m$ may differ from our estimate $\Delta
m_{\rm est}$. In the limit of a large number of measurements the phase
distribution is then governed by the likelihood function
\begin{equation}
F(\phi|\theta;\Delta m_{\rm est},\Delta m)={\mathcal N}^{-1} \prod_m P_{\hat\rho(\Delta m_{\rm est})}(m|\phi)^{P_{\hat \rho(\Delta m)}(m|\theta)}.
\end{equation} 
That function is shown in figure (\ref{Fphidmest}) for $J=20$, $\Delta m=2$,
and $\theta=0$. The figure illustrates two typical properties of
$F(\phi|\theta=0;\Delta m_{\rm est.},\Delta m)$. If the estimate of the fluctuations is too optimistic, i.e. if $\Delta m_{\rm est}<\Delta m$,
the likelihood is peaked at a non-zero phase
shift and it has a minimum at zero phase shift. For such estimates the phase distribution is asymptotically inaccurate. This is because for a too
optimistic estimate of $\Delta m$ one observes non-zero number differences at
the output too often to be brought about by a non-zero phase shift. One thus
wrongly concludes that the phase shift was non-zero.

For conservative estimates $\Delta m_{\rm est}>\Delta m$ on the other hand the
likelihood function of the phase is peaked at zero but it is broadened compared
to the likelihood function for perfect knowledge of the number fluctuations,
$\Delta m_{\rm est}=\Delta m$. As expected, the best phase resolution is
obtained for $\Delta m_{\rm est}=\Delta m$.

\section{\label{conclusion}Conclusion}

Interferometry with balanced Fock states is robust against imperfections in the
state preparation. The phase resolution is insensitive to noise in the total
number of particles. For number difference uncertainties smaller than one the
phase resolution is essentially Heisenberg limited. For larger state
preparation imperfections the phase resolution deteriorates linearly with the
number difference uncertainty.

In future work we plan to study the robustness of interferometry with balanced
Fock states against dephasing and decoherence inside the interferometer as well
as the effect of non-ideal detectors.

\ack

We gratefully acknowledge stimulating discussions with H. Uys and R. Pepino.
This work has been supported by DAAD, DFG, and NSF.

\section*{References}

\bibliography{mybibliography}

\end{document}